\documentclass[twocolumn]{aastex61}

\usepackage{graphicx}
\usepackage{natbib}

\received{July 1, 2016}
\revised{September 27, 2016}
\accepted{\today}
\submitjournal{ApJ}

\shorttitle{SOLIS presentation}
\shortauthors{Ceccarelli et al.}

\begin{document}

\title{Seeds Of Life In Space (SOLIS): The organic composition diversity
  at 300--1000 au scale in Solar-type star forming regions 
  \footnote{Based on observations carried out under project number L15AA
    with the IRAM NOEMA Interferometer.  IRAM is supported by
    INSU/CNRS (France), MPG (Germany) and IGN (Spain).}}  
\correspondingauthor{C. Ceccarelli \& P. Caselli}
\email{cecilia.ceccarelli@univ-grenoble-alpes.fr, caselli@mpe.mpg.de}

\author{C. Ceccarelli}
\affil{Univ. Grenoble Alpes, CNRS, IPAG, F-38000 Grenoble, France}
\author{P. Caselli}
\affil{Max-Planck-Institut für extraterrestrische Physik, Giessenbachstrasse 1, 85748  Garching,  Germany}
\author{F. Fontani}
\affil{INAF-Osservatorio Astrofisico di Arcetri, Largo E. Fermi 5, I-50125, Florence, Italy}
\author{R. Neri}
\affil{Institut de Radioastronomie Millim\'etrique, 300 rue de la Piscine, 38406, Saint-Martin d'H\`eres, France}
\author{A. L\'opez-Sepulcre}
\affil{Institut de Radioastronomie Millim\'etrique, 300 rue de la Piscine, 38406, Saint-Martin d'H\`eres, France}
\affil{Univ. Grenoble Alpes, CNRS, IPAG, F-38000 Grenoble, France}
\author{C. Codella}
\affil{INAF-Osservatorio Astrofisico di Arcetri, Largo E. Fermi 5, I-50125, Florence, Italy}
\author{S. Feng}
\affil{Max-Planck-Institut für extraterrestrische Physik, Giessenbachstrasse 1, 85748  Garching,  Germany}
\author{I. Jim\'enez-Serra}
\affil{Department of Physics and Astronomy, University College London, Gower St., London, WC1E 6BT, UK}
\affil{School of Physics and Astronomy, Queen Mary University of London, 327 Mile End Road, London, E1 4NS}
\author{B. Lefloch}
\affil{Univ. Grenoble Alpes, CNRS, IPAG, F-38000 Grenoble, France}
\author{J. E. Pineda}
\affil{Max-Planck-Institut für extraterrestrische Physik, Giessenbachstrasse 1, 85748  Garching,  Germany}
\author{C. Vastel}
\affil{Universit\'e de Toulouse, UPS-OMP, IRAP, Toulouse, France}
\affil{CNRS, IRAP, 9 Av. Colonel Roche, BP 44346, F-31028 Toulouse Cedex 4, France}
\author{F. Alves}
\affil{Max-Planck-Institut für extraterrestrische Physik, Giessenbachstrasse 1, 85748  Garching,  Germany}
\author{R. Bachiller}
\affil{Observatorio Astron\'omico Nacional (OAN, IGN), Calle Alfonso XII, 3, 28014 Madrid, Spain}
\author{N. Balucani}
\affil{Dipartimento di Chimica, Biologia e Biotecnologie, Universit\`a di Perugia, Via Elce di Sotto 8, I-06123 Perugia, Italy}
\affil{Univ. Grenoble Alpes, CNRS, IPAG, F-38000 Grenoble, France}
\affil{INAF-Osservatorio Astrofisico di Arcetri, Largo E. Fermi 5, I-50125, Florence, Italy}
\author{E. Bianchi}
\affil{Univ. Grenoble Alpes, CNRS, IPAG, F-38000 Grenoble, France}
\affil{Dipartimento di Fisica e Astronomia, Universit\'a degli Studi
  di Firenze, Italy}
\author{L. Bizzocchi}
\affil{Max-Planck-Institut für extraterrestrische Physik, Giessenbachstrasse 1, 85748  Garching,  Germany}
\author{S. Bottinelli}
\affil{Universit\'e de Toulouse, UPS-OMP, IRAP, Toulouse, France}
\affil{CNRS, IRAP, 9 Av. Colonel Roche, BP 44346, F-31028 Toulouse Cedex 4, France}
\author{E. Caux}
\affil{Universit\'e de Toulouse, UPS-OMP, IRAP, Toulouse, France}
\affil{CNRS, IRAP, 9 Av. Colonel Roche, BP 44346, F-31028 Toulouse Cedex 4, France}
\author{A. Chac\'on-Tanarro}
\affil{Max-Planck-Institut für extraterrestrische Physik, Giessenbachstrasse 1, 85748  Garching,  Germany}
\author{R. Choudhury}
\affil{Max-Planck-Institut für extraterrestrische Physik, Giessenbachstrasse 1, 85748  Garching,  Germany}
\author{A. Coutens}
\affil{Department of Physics and Astronomy, University College London, Gower St., London, WC1E 6BT, UK}
\author{F. Dulieu}
\affil{LERMA, Universit\'e de Cergy-Pontoise, 95000 Cergy Pontoise Cedex, France}
\author{C. Favre}
\affil{Univ. Grenoble Alpes, CNRS, IPAG, F-38000 Grenoble, France}
\affil{INAF-Osservatorio Astrofisico di Arcetri, Largo E. Fermi 5, I-50125, Florence, Italy}
\author{P. Hily-Blant}
\affil{Univ. Grenoble Alpes, CNRS, IPAG, F-38000 Grenoble, France}
\author{ J. Holdship}
\affil{Department of Physics and Astronomy, University College London, Gower St., London, WC1E 6BT, UK}
\author{C. Kahane}
\affil{Univ. Grenoble Alpes, CNRS, IPAG, F-38000 Grenoble, France}
\author{A. Jaber Al-Edhari}
\affil{Univ. Grenoble Alpes, CNRS, IPAG, F-38000 Grenoble, France}
\affil{University of AL-Muthanna, College of Science, Physics Department, AL-Muthanna, Iraq}
\author{J. Laas}
\affil{Max-Planck-Institut für extraterrestrische Physik, Giessenbachstrasse 1, 85748  Garching,  Germany}
\author{J. Ospina}
\affil{Univ. Grenoble Alpes, CNRS, IPAG, F-38000 Grenoble, France}
\author{Y. Oya}
\affil{Department of Physics, The University of Tokyo, Bunkyo-ku, Tokyo 113-0033, Japan}
\author{L. Podio}
\affil{INAF-Osservatorio Astrofisico di Arcetri, Largo E. Fermi 5, I-50125, Florence, Italy}
\author{A. Pon}
\affil{Max-Planck-Institut für extraterrestrische Physik, Giessenbachstrasse 1, 85748  Garching,  Germany}
\affil{Department of Physics and Astronomy, The University of Western Ontario, 1151 Richmond Street, London, N6A 3K7, Canada}
\author{A. Punanova}
\affil{Max-Planck-Institut für extraterrestrische Physik, Giessenbachstrasse 1, 85748  Garching,  Germany}
\author{D. Quenard}
\affil{Department of Physics and Astronomy, University College London, Gower St., London, WC1E 6BT, UK}
\affil{School of Physics and Astronomy, Queen Mary University of London, 327 Mile End Road, London, E1 4NS}
\author{A. Rimola}
\affil{Departament de Qu\'imica, Universitat Aut\`onoma de Barcelona, E-08193 Bellaterra, Spain}
\author{N. Sakai}
\affil{The Institute of Physical and Chemical Research (RIKEN), 2-1, Hirosawa, Wako-shi, Saitama 351-0198, Japan}
\author{I.R. Sims}
\affil{Institut de Physique de Rennes, UMR CNRS 6251, Université de Rennes 1, 263 Avenue du Général Leclerc, F-35042 Rennes Cedex, France}
\author{S. Spezzano}
\affil{Max-Planck-Institut für extraterrestrische Physik, Giessenbachstrasse 1, 85748  Garching,  Germany}
\author{V. Taquet}
\affil{INAF-Osservatorio Astrofisico di Arcetri, Largo E. Fermi 5, I-50125, Florence, Italy}
\author{L. Testi}
\affil{INAF-Osservatorio Astrofisico di Arcetri, Largo E. Fermi 5, I-50125, Florence, Italy}
\affil{European Southern Observatory, Karl-Schwarzschild-Str. 2, 85748 Garching bei M\"{u}nchen, Germany}
\author{P. Theul\'e}
\affil{Aix-Marseille Universit\'e, PIIM UMR-CNRS 7345, 13397 Marseille, France}
\author{P. Ugliengo}
\affil{Dipartimento di Chimica and NIS Centre, Universit\`a degli Studi di Torino, Via P. Giuria 7, I-10125 Torino, Italy}
\author{A.I. Vasyunin}
\affil{Max-Planck-Institut für extraterrestrische Physik, Giessenbachstrasse 1, 85748  Garching,  Germany}
\affil{Ural Federal University, Ekaterinburg, Russia}
\author{S. Viti}
\affil{Department of Physics and Astronomy, University College London, Gower St., London, WC1E 6BT, UK}
\author{L. Wiesenfeld}
\affil{Univ. Grenoble Alpes, CNRS, IPAG, F-38000 Grenoble, France}
\author{S. Yamamoto}
\affil{Department of Physics, The University of Tokyo, Bunkyo-ku, Tokyo 113-0033, Japan}
\date{Received date; accepted date}

\begin{abstract} 
  Complex organic molecules have been observed for decades in the
  interstellar medium. Some of them might be considered as small
  bricks of the macromolecules at the base of terrestrial life. It is
  hence particularly important to understand organic chemistry in
  Solar-like star forming regions. In this article, we present a new
  observational project: SOLIS (Seeds Of Life In Space).  This is a
  Large Project at the IRAM-NOEMA interferometer, and its scope is to
  image the emission of several crucial organic molecules in a sample
  of Solar-like star forming regions in different evolutionary stages
  and environments.  Here, we report the first SOLIS results, obtained
  from analysing the spectra of different regions of the Class 0
  source NGC1333-IRAS4A, the protocluster OMC-2 FIR4, and the shock
  site L1157-B1. The different regions were identified based on the
  images of formamide (NH$_2$CHO) and cyanodiacetylene (HC$_5$N)
  lines. We discuss the observed large diversity in the molecular and
  organic content, both on large (3000--10000 au) and relatively small
  (300--1000 au) scales. Finally, we derive upper limits to the
  methoxy fractional abundance in the three observed regions of the
  same order of magnitude of that measured in few cold prestellar
  objects, namely $\sim 10^{-12}$--$10^{-11}$ with respect to H$_2$
  molecules.
 \end{abstract}

\keywords{ISM: abundances  ---  ISM: clouds -- ISM: molecules -- Radio lines: ISM}


\section{Introduction}
\label{intro}

From a “simple” chemical point of view, all terrestrial living
organisms, from microbes to humans, are made up of the same basic
components: amino acids, fatty acids, sugars, nucleobases, etc. In
total, we are referring to about 50 “small” molecules containing less
than 100 atoms of carbon, with hydrogen, oxygen, nitrogen and other
elements in smaller quantities: terrestrial life is based on organic
chemistry.  Of course, this is not by chance, but due to the
electronic structure of C atoms and their abundance (not locked in
refractory/rocky material). Thus, it does not come as a surprise that
the largest molecules detected in the ISM, and those having more than
six atoms, contain carbon {\it
  (http://www.astro.uni-koeln.de/cdms/molecules}; M\"uller et al.
2001). Moreover, species like formamide (NH$_2$CHO), believed to be a
crucial molecule in the synthesis of metabolic and genetic species in
modern versions of the Urey-Miller experiment (Saladino et al. 2012),
are easily found in regions forming Solar-like stars (Kahane et al.
2013; Mendoza et al. 2014; L\'opez-Sepulcre et al. 2015) as well as in
external galaxies (M\"uller et al. 2013). One step farther, amino
acids are found in cometary and meteoritic material (Elsila et al.
2009; Altwegg et al. 2016; Pizzarello et al. 1991). This evidence led
the Nobel laureate C. De Duve to affirm: “the chemical seeds of life
are universal” and “life is an obligatory manifestation of matter,
written into the fabric of the Universe” (De Duve 2005, 2011).

One of the remarkable discoveries of modern Astronomy is that
relatively complex organic molecules are present in the interstellar
medium (ISM) (e.g. van Dischoeck \& Herbst 2009). In the following, we
will refer to C-bearing molecules containing at least six atoms as
iCOMs, for ``interstellar Complex Organic Molecules''\footnote{Please
  note that we added ``i'' to the commonly used COMs acronym in order
  to be clear that these molecules are only complex in the
  interstellar context contrarily to what chemists would consider
    complex in the terrestrial context.}. They may be seen as the
smallest bricks from which the larger ``chemical seeds of life''
mentioned by De Duve are formed.  Although iCOMs have been observed
for decades in massive star formation regions (e.g. Rubin et
al. 1971), the detection of these special molecules in regions that
will eventually form Solar-like planetary systems came much later
(Ceccarelli et al. 2000b; Cazaux et al. 2003). This was a crucial
discovery that set a direct link between organic chemistry in the ISM
and in the Solar System, and provided additional and crucial ground to
the ``universal chemical seeds of life'' hypothesis.

Yet, despite the great importance of iCOMs, their routes of formation
and destruction are still largely debated
(e.g. Herbst \& van Dischoeck 2009; Caselli \& Ceccarelli 2012; Herbst
\& Vasyunin 2013; Balucani et al. 2015; Enrique-Romero et al. 2016;
Vasyunin et al. 2017).  The reasons for this impasse are multiple. One
of them is the lack of systematic observational studies which would
put strong constraints on theory, when observations are compared to
models. In particular, the vast majority of the 43 so far detected
iCOMs are only observed towards SgrB2 (e.g.  Belloche et al. 2017).
In Solar-like star forming regions, only 18 iCOMs have been detected
so far (Table \ref{tab:iCOMs}).
\begin{table}
  \caption{List of the iCOMs detected in Solar-like star forming
    regions. Last column provides representative relevant references.}\label{tab:iCOMs}
  \begin{tabular}{lll}
    \hline \hline
    Molecule name & Formula & References\\
  \hline 
    Methanol & CH$_3$OH & 1, 2\\
    Methanethiol & CH$_3$SH & 3 \\
    Methyl cyanide & CH$_3$CN & 4\\
    Formamide & NH$_2$CHO & 5\\
    Propyne & CH$_3$CCH & 6, 7\\
    Ethylene oxide & c-C$_2$H$_4$O & 8\\
    Acetaldehyde & CH$_3$CHO & 6, 9, 10\\
    Methyl isocyanate & CH$_3$NCO & 11, 12\\
    Cyanoacetylene & HC$_5$N & 13\\
    Methyl formate & HCOOCH$_3$ & 14, 15, 16\\
    Glycol aldehyde & HCO(CH$_2$)OH & 17, 18, 19\\
    Acetic acid & CH$_3$COOH & 20\\
    Dimethyl ether & CH$_3$OCH$_3$ & 14, 13, 16\\
    Ethanol & CH$_3$CH$_2$OH & 21, 18, 16 \\
    Ethyl cyanide & CH$_3$CH$_2$CN & 14, 15, 22\\
    Acetone & CH$_3$COCH$_3$ & 8\\
    Propanal & CH$_3$CH$_2$CHO & 8\\
    Ethylene glycol & (CH$_2$OH)$_2$ & 17\\
    \hline
  \end{tabular}\\
  {\small References: 1- Bizzocchi et al. 2014; 2- Maret et al. 2005;
    3- Majumdar et al. 2016; 4- van Dishoeck et al. 1995;
    5- L\'opez-Sepulcre et al. 2015; 6- Vastel et al. 2014; 7- Caux et
    al. 2011; 8- Lykke et al. 2017; 9- Jaber et al. 2014; 10-
    Codella et al. 2015; 11- Ligterink et al. 2017; 
    12- Martin-Dom\'enec et al. 2017; 13- Jaber Al-Edhari et al. 2017;
    14- Jim\'enez-Serra et al. 2016; 15- Cazaux et al. 2003; 16-
    Lefloch et al. 2017; 17- J\o rgensen et al. 2012; 18- Taquet et
    al. 2015; 19- Coutens et al. 2015; 20- J\o rgensen et al. 2016; 21-
    Bisschop et al. 2008; 22- L\'opez-Sepulcre et al. 2017.
}
\end{table}
Probably even more important for building up a theory, most of these
iCOMs are observed in very few sources. The most studied ones are the
hot corino sources, of which the prototype and best studied one is
IRAS16293-2422 (e.g. van Dishoeck et al. 1995; Ceccarelli et
al. 2000a; Jaber et al. 2014; J\o rgensen et al. 2016).
iCOMs are present in large quantities also in the shocks created by
the outflows from Solar-like protostars, but also here studies exist
only towards a very limited number of sources (e.g.  Requena-Torres et
al. 2007; Arce et al. 2008; Codella et al. 2015; Lefloch et al.
2017a). Interestingly, iCOMs are present in the lukewarm clouds of the
Galactic Center (Requena-Torres et al. 2006), likely due to the
widespread shocks in the region.
 Finally and surprisingly, iCOMs have also been detected in very few
 of the coldest known sources, the prestellar cores (\"Oberg et
 al. 2010; Bacmann et al. 2012; Cernicharo et al. 2012; Vastel et
 al. 2014; Jim\'enez-Serra et al.  2016).
In summary, the observational framework is too sparse to provide
strong constraints to theory.

In this respect, the case of methyl formate (MF) dimethyl ether (DME)
provides an illustrative example of why observations towards a larger
number of objects and with a better spatial resolution are needed to
firmly establish how these species are formed. The available
observations, cited above, show that MF and DME are present in hot and
cold environments, and that their abundances are correlated
(e.g. Jaber et al. 2014). Early models claimed that both species are
synthesised on the grain surfaces, thanks to reactions between
radicals created by the UV illumination of the grain mantles during
the cold ($\leq20$ K) phase (e.g. Garrod et al. 2008 and the so-called
warm-up phase models). However, radicals become mobile only at dust
temperatures larger than about 25-30 K, so that in cold environments
the process would be inhibited. In addition, the liberation of MF and
DME from the grain surfaces at low temperature would be problematic
(e.g. Minissale et al. 2016). Alternatively, it has been suggested
that MF and DME are synthesised in the gas-phase from the methanol,
formed on the grain surface by the hydrogenation of CO and UV
photodesorbed (e.g. Balucani et al. 2015).  Spatially resolved maps of
the two species, plus that of methanol, in different regions where the
thermal structure is well known and with different UV radiation field
would allow to distinguish which process dominates in what
environment.

In order  to attack this  problem, we started  a Large Program  at the
NOEMA interferometer called SOLIS ({\it  Seeds Of Life In Space}). The
immediate  goal of  SOLIS  is  to provide  an  homogeneous dataset  of
observations  of   five  crucial  iCOMs   in  half  a   dozen  targets
representative  of  Solar-like  systems in  their  first  evolutionary
stages.  The ultimate  goal is  to have  a more  complete view  of how
organic  chemistry starts  and  develops during  the  first stages  of
evolution  of Solar-like  systems.  The present  article presents  the
overall SOLIS project and some results based on the first observations
around 82 GHz, with a  moderate spatial resolution. A series of
accompanying  articles  provides   additional  and  detailed  analysis
towards single  SOLIS targets: L1157-B1  by Codella et al.  (2017) and
Feng et al. (in prep.); OMC-2 FIR4  by Fontani et al. (2017), Favre et
al. (in prep.) and Neri et  al. (in prep.); IRAS4A by L\'opez-Sepulcre
et al. (in  prep.); L1544 by Punanova et al.  (2017). Forthcoming
  articles  will present  the remaining  observations obtained also
  with higher spatial resolution.

The present article is organised as follows. We present the targeted
iCOMs and their transitions in \S \ref{sec:projet-presentation}
together with the targeted sources, and the motivations for the
choices. In \S \ref{sec:observations-results} we present the results
of the first observations, obtained with the same frequency
setting, which permits a first analysis of the similarity and
differences of the observed sources, as discussed in \S
\ref{sec:discussion}. A final section, \S \ref{sec:conclusions},
concludes the article.


\section{Project presentation}\label{sec:projet-presentation}

The goal of the SOLIS project is to build an homogeneous dataset of
observations from which we can constrain the routes of formation and
destruction of a selected number of iCOMs during the earliest phases
of formation of a Solar-like system. To this end, we selected a sample
of sources representative of these first phases, described in \S
\ref{sec:source-sample}, and obtained high spatial resolution images
of selected lines of a sample of molecules, described in \S
\ref{sec:target-species-lines}, crucial to understand the major routes
of formation and destruction of iCOMs.

\subsection{Source sample}\label{sec:source-sample}
SOLIS aims to cover the first evolutionary phases of the Solar-type star
formation process, by observing several sources even though the sample
will inevitably be incomplete. The sources were selected based on the
results of two previous Large Programs: CHESS ({\it Chemical HErschel
  Surveys of Star forming regions}; Ceccarelli et al. 2010) and ASAI
({\it Astrochemical Surveys At Iram}; Lefloch et al. 2017b). Both
programs consist of single-dish unbiased spectral surveys of
representative star forming regions: CHESS covers the 500--1900 GHz
band with Herschel/HIFI, whereas ASAI covers the 3, 2 and 1.3 mm bands
observable with the IRAM-30m telescope. The SOLIS selected sources are
listed in Table \ref{tab:sources}.
\begin{table*}
  \caption{List of the SOLIS targets. The different columns
    report the name of each source, type (see text for more details),
    luminosity,  distance, coordinates of the center of the
    observations and local standard of rest velocity. The references
    to the source type, luminosity 
    and distance are reported in the text. The last two columns report
    whether each source is a target of the ASAI and CHESS Large
    Programs (Lefloch et al. 2017b; Ceccarelli et al. 2010).}\label{tab:sources}
  \begin{tabular}{l|ccc|ccc|cc}
    \hline \hline
    Source & Type & Lum         & Dist & RA   & DEC & V$_{LSR}$ & ASAI & CHESS\\
              &         & (L$_\odot$) & (pc) &        &       &               & & \\
    \hline 
    L1544        & prestellar core & -     & 140  & 05:04:12.2 & 25:10:42.8 & 7.3 & Y & Y\\
    L1521F      & VeLLO               & 0.1  & 140  & 04:28:39.0 & 26:51:35.6 & 6.4 & N & N\\
    NGC1333-IRAS4A   & Class 0 & 10   & 260  & 03:29:10.5 & 31:13:30.9 & 6.5 & Y & Y\\
    CepE                    & Class 0 & 100  & 730  & 23:03:13.0 & 61:42:21.0 & -10.9& Y & N\\
    NGC1333-SVS13A & Class I  & 34    & 260  & 03:29:03.7 & 31:16:03.8 & 8.5& Y & N\\
    OMC-2 FIR4 & proto-cluster & $\leq$1000 & 420 & 05:35:27.0 & -05:09:56.8 & 11.4 & Y & Y \\
    L1157-B1 & molecular shock & -     & 250  & 20:39:10.2 & 68:01:10.5 & -2.6 & Y & Y\\
    \hline
  \end{tabular}
\end{table*}
They are all well-known sources representative of different
evolutionary phases: the prestellar core L1544, the VeLLO (Very Low
Luminosity Object) L1521F, the low mass Class 0 source NGC1333-IRAS4A,
the intermediate mass Class 0 source CepE, the low mass Class I source
NGC1333-SVS13A, the protocluster OMC-2 FIR4, and the molecular shock
L1157-B1. In the following we provide a brief description of each
source.

\subsubsection{L1544} 
L1544 is a prestellar core in the Taurus complex, at a distance of 140
pc (Kenyon et al. 1994; Schlafly et al. 2014). This prestellar core
has been studied extensively and its chemical and physical structure
is well-known. The gas temperature reaches a minimum of 6 K toward the
central 2000 au and a maximum of 13 K at 15,000 au (Crapsi et
al. 2007). Caselli et al. (1999) measured large amounts of CO
freeze-out within the central few thousands au and, within the
CO-depleted zone, deuterated molecules thrive (Caselli et al.  2002;
Caselli et al. 2003; Bacmann et al. 2003; Vastel et al. 2006; Crapsi
et al. 2007). The ground state transition of ortho-H$_2$O reveals an
inverse P-Cygni profile which can be reproduced with a simple
chemical-dynamical model describing the quasi-equilibrium contraction
of a Bonnor-Ebert sphere (Caselli et al. 2012; Keto et al. 2015;
Quenard et al. 2016). L1544 is, hence, a prestellar core on the verge
of star formation. Thus, its study will provide us with crucial
information about the chemical processes before the switch-on of the
protostar.  Several iCOMs are detected towards this core (Bizzocchi et
al. 2014; Vastel et al. 2014; Jim\'enez-Serra et al. 2016). NOEMA
observations are centered on the position of the CH$_3$OH peak
(Bizzocchi et al.  2014), where iCOMs are also expected to peak
(Vastel et al. 2014; Jimenez-Serra et al. 2016), about 5000 au away
from the dust peak, as confirmed by the theoretical work of Vasyunin
et al. (2017).

\subsubsection{L1521F} 
L1521F is another dense core in the Taurus molecular cloud. It was
studied by Crapsi et al. (2004), who found a striking similarity with
L1544, with the exception of a factor of 2 lower deuterium
fraction and a peculiar velocity field. Based on these results,
Crapsi et al. (2004) defined L1521F "a highly evolved starless
core". In fact, two years later, a faint infrared source was
discovered at its dust peak by the {\em Spitzer Space Telescope}
(Bourke et al. 2006). This source belongs to the group of VeLLOs, i.e.
objects with a luminosity lower than about 0.1 L$_\sun$, embedded in
dense cores (Kauffmann et al. 2005; Dunham et al. 2006). More
recently, the central regions of L1521F were studied with ALMA: Tokuda
et al. (2014) imaged dust continuum emission and simple molecules
(HCO$^+$, HCN, CS), finding large central densities and sub-structure
with an arc-like morphology. This structure resembles the
hydrodynamical simulations of infalling material around protostars
carried out by Matsumoto et al. (2015). Tokuda et al. (2016) studied
in detail the structure of the L1521F core, from tens to $\sim$10000
au, revealing an inner region with a flat density structure where an
unresolved protostellar source is next to a starless high-density
core. In summary, L1521F represents the next evolutionary stage of a
prestellar core such as L1544.

\subsubsection{NGC1333-IRAS4A} 
NGC1333-IRAS4A is located in the Perseus NGC1333 region at a distance
of (260$\pm$26) pc (Hirota et al. 2008; Schlafly et al. 2014). It was the
second discovered hot corino (Bottinelli et al. 2004), namely hot
($\geq 100$ K), dense ($\geq 10^7$ cm$^{-3}$) and compact ($\leq$ 100
au in radius) regions enriched of iCOMs (Ceccarelli et al. 2007;
Caselli \& Ceccarelli 2012).  Interferometric observations showed that
it is a binary system composed of two Class 0 objects separated by
$\sim$1$\farcs$8 (423 au): IRAS4-A1 and IRAS4-A2 (e.g. Looney et
al. 2000; Santangelo et al.  2015; Tobin et al. 2016). The overall
bolometric luminosity of IRAS4A and its total envelope mass are
estimated to be 9.1 L$_\sun$ and 5.6 M$_\sun$, respectively
(Kristensen et al. 2012; Karska et al. 2013; Tobin et al. 2016). While
IRAS4-A1 is more than three times brighter at 1.3 mm than its
companion, only IRAS4-A2 is associated with a hot corino (Taquet et
al. 2015; Coutens et al. 2015; De Simone et al. 2017; Lop\'ez-Sepulcre
et al. 2017). In addition, IRAS4-A1 drives a fast collimated jet
associated with bright H$_2$ emission, whereas IRAS4-A2 powers a
slower and precessing jet (Santangelo et al. 2015).  These findings
suggest that the two Class 0 sources are likely in different
evolutionary stages, although it is not possible so far to firmly
conclude whether this is due to evolution or rather accretion
luminosity effects/bursts (De Simone et al. 2017; Lop\'ez-Sepulcre et
al.  2017). The compact jets from IRAS4-A1 and IRAS4-A2 create a
shock site $\sim12''$ south, where SOLIS observations show emission
from iCOMs, for the first time (see \S \ref{sec:results}).

\subsubsection{CepE} 
CepE is located in the second most massive molecular
cloud of the Cepheus OB3 association at a distance of 730 pc (Sargent,
1977; Few \& Cohen 1983). It is an isolated Class 0 source of
intermediate mass associated with the infrared source IRAS 23011+6126.
Its luminosity is $\sim100$ L$_\sun$ and its mass is $35$ M$_\sun$
(Lefloch et al. 1996; Crimier et al. 2009). CepE drives an
exceptionally powerful outflow, the southern lobe of which is
terminated by the Herbig Haro object HH377 (Lefloch et al. 1996,
2015). Studies of the millimeter CO and the near-infrared H$_2$ and
[FeII] lines revealed the presence of several outflows (Ladd \& Hodapp
1997; Eislöffel et al. 1996). Continuum and line observations
obtained with the IRAM- PdBI (Plateau de Bure Interferometer) at 1$''$
scale confirm the presence of a binary protostellar system, with each
component driving a high-velocity jet detected in lines of
H$_2$, CO and SiO (Ospina-Zamudio et al.  in prep). The dynamical
timescale of these ejections is short, typically $1000$ yr. The same
observations also show evidence for the presence of a hot corino in
one of the two components, so that this is a good target to understand
the influence of luminosity on the formation and destruction of iCOMs.

\subsubsection{NGC1333-SVS13A} 
NGC1333-SVS13A is located in the Perseus NGC1333 region at a distance
of (260$\pm$26) pc (Hirota et al. 2008; Schlafly et al. 2014), close
to the other SOLIS source NGC1333-IRAS4A. SVS13A is part of a small
cluster of protostars, of which the two brightest sources, labelled A
and B, are separated by $\sim 15''$ (e.g. Chini et al. 1997; Bachiller
et al. 1998; Looney et al. 2000; Chen et al. 2009; Tobin et
al. 2016). SVS13A is relatively evolved, as it possesses a ratio
L$_{\rm submm}$/L$_{\rm bol} \sim 0.08$, typical of Class I
sources. Its luminosity is $\sim 33$ L$_\sun$ (Tobin et
al. 2016). SVS13A drives the famous chain of Herbig-Haro (HH) objects
7-11 (Reipurth et al. 1993) and is associated with an extended outflow
($>$0.07 pc: Lefloch et al. 1998, Codella et al.  1999). For
comparison, SVS13B is a Class 0 protostar with L$_{\rm bol} \sim 1$
L$_\sun$ (Tobin et al. 2016; De Simone et al.  2017) that drives a
well-collimated SiO jet (Bachiller et al. 1998).  Recent observations
revealed the presence of a compact ($\sim$50 au), hot ($\sim 200$ K)
and dense ($\geq10^7$ cm$^{-3}$) region in SVS13A, hinting at the
presence of a hot corino region, the first ever detected in a Class I
source (Codella et al. 2016; De Simone et al.  2017). More recently,
Bianchi et al. (2017) measured a deuterium fraction in SVS13A smaller
than in Solar-like prestellar cores and Class 0 sources (their Figure
9), showing, for the first time, that there may be a chemical
evolution between Class 0 and I sources.  Therefore, SVS13A is a
perfect target to understand the evolutionary effects in the iCOMs
abundances, namely formation and destruction routes.

\subsubsection{OMC--2 FIR4} 
OMC--2 FIR4 belongs to the Orion Molecular Complex, at a distance of
420 pc (Hirota et al. 2007; Schlafly et al. 2014). The total FIR4 mass
and luminosity are around 30 M$_\sun$ (Mezger et al. 1990; Crimier et
al. 2009) and less than 1000 L$_\sun$ (Crimier et al. 2009; Furlan et
al. 2014), respectively.  Interferometric observations showed that
FIR4 contains several protostars and clumps, even though their number
is uncertain (Shimajiri et al. 2008; L\'opez-Sepulcre et al. 2013;
Gonzalez-Garcia et al. 2016).Our new SOLIS maps detect at least eight
sources, of which only one has a FIR counterpart so that the others
are, likely, very young Class 0 or prestellar objects (Neri et al. in
prep.). Therefore, FIR4 is certainly a young protocluster. The nature
of the hosted objects is practically unknown: their mass, temperature,
chemical composition and dynamical status. However, what makes this
protocluster particularly interesting and the reason why it is a SOLIS
target is that FIR4 is, at present, the best known analogue of the Sun
progenitor for two reasons. The first one is that the Sun was born in
a cluster of stars and not in an isolated clump (Adams 2010). The
second reason is that the Solar System experienced a large irradiation
of energetic ($\geq10$ MeV) particles during its first phases of
formation (Gounelle et al. 2006). A similar dose of energetic
particles is present somewhere in or close to FIR4, producing an
enhanced degree of gas ionisation (Ceccarelli et al. 2014; Fontani et
al. 2017). Finally, close to FIR4 lie two additional sources, FIR3 and
FIR5, about 25$''$ north and south, respectively. Not much was known
about these two sources before our SOLIS observations. Shimajiri et
al. (2015) obtained an unbiased spectral survey at 3 and 0.8 mm
towards FIR3, with single-dish telescopes and corresponding angular
resolution of about 20$"$. Based on those observations, Shimajiri et
al. claim that the observed line emission is dominated by the shocked
gas from the outflow rising from FIR3 and hitting FIR4 rather than a
hot core/corino sources.

\subsubsection{L1157-B1} 
This famous shocked region is associated with
the Class 0 protostar L1157-mm, at a distance of 250 pc (Looney et
al. 2007).  L1157-mm has a luminosity of $\simeq$ 3 $L_\sun$ (Tobin et
al.  2010), and drives an episodic and precessing jet (Gueth et
al. 1996, 1998; Podio et al. 2016).  In the last $\leq 2000$ yr, the
jet accelerated and entrained material, creating a spectacular bipolar
outflow and opening cavities along the path. The south and
blue-shifted lobe of the outflow is formed by two cavities, the
respective apexes of which are known as B1 and B2 (e.g. Gueth et
al. 1998). The latter is more distant from L1157-mm and, hence, older
than B1. The SOLIS target is B1, the age of which is estimated to be
less than 2000 yr (Podio et al. 2016). This low age and the extent of
the shock allow us to probe chemical composition variations along the
shock and, consequently, to test the chemistry where the dust icy
mantles are recently released in the gas phase via sputtering. In the
context of the SOLIS goal, B1 provides us with the possibility to test
the different theories of iCOMs formation. Indeed, the different
ejection episodes caused several shocks where the jet impacted the
cavities walls. In these sites, the abundance of a large number of
species is dramatically enhanced: from diatomic, such as SiO and CS,
to more complex molecules, such as CH$_3$CN and CH$_3$CHO (see
e.g. Bachiller et al. 2001; Benedettini et al. 2013; Codella et
al. 2015). Lately, several iCOMs were detected in B1 with abundances
as large as or more than in hot corinos (Lefloch et
al. 2017a). However, just a few iCOMs (methanol, methyl cyanide and
acetaldehyde) have been imaged with interferometers so far (Codella et
al. 2007, 2015; Fontani et al. 2014). These images show a
inhomogeneous distribution of the molecules, associated with the
different ejection episodes. The comparison of the distribution of
acetaldehyde and formamide obtained with the new SOLIS observations
provides us with very stringent constraints on the formation route of
formamide (Codella et al. 2017).

\subsection{Target species and lines}\label{sec:target-species-lines}
In the zoo of the detected iCOMs, we selected five species to observe
with high spectral resolution, as summarised in Table
\ref{tab:SOLIS-molecules}. These species were chosen because they are
known to be abundant in warm ($\geq 50$ K) and, sometimes, cold
($\leq 20$ K) environments, and have the potential to
discriminate between the principal chemical mechanisms at work,
specifically: (i) grain surface versus gas phase chemistry, (ii) the
exact formation and destruction routes, e.g. asserting whether a class
of reactions, such as gas radiative associations or surface
photolysis, plays an important role, and (iii) the desorption yield
and efficiency. Table \ref{tab:SOLIS-molecules} provides a very short
summary of what is known or hypothesised about their formation routes.

To give a practical example, two major routes of formation of
formamide are discussed in the literature: (1) on the grain surfaces,
either via reactions occurring between radicals when the grain
temperature allows them to be mobile (at $\geq30$ K; e.g. Garrod \&
Herbst 2006; Garrod et al. 2008) or thanks to UV/particle ice
irradiation (Bennett \& Kaiser 2007; Jones et al.  2011; Kanuchova et
al. 2016; Fedoseev et al. 2016); (2) in the gas phase from the
reaction of formaldehyde (H$_2$CO) and amidogen (NH$_2$) (Kahane et
al. 2013; Barone et al. 2015; Vazart et al. 2016; Skouteris et
al. 2017). Formaldehyde and amidogen are the products of hydrogenation
on the grain surfaces of CO and N, respectively, occurring during the
cold prestellar phase (e.g. Vasyunin et al. 2017). The SOLIS
observations of the distribution of the formamide abundance compared
to other iCOMs can distinguish between the two theories, as they have
different formation and destruction timescales, as we show in Codella
et al. (2017).

More generally, abundance maps of iCOMs in different environments,
cold and warm, also help to find which theory is the correct one and
to constrain the underlying chemistry, including the desorption
mechanism in different zones: interstellar UV photons versus photons
generated by cosmic ray impacts with H$_2$ molecules, chemical
reactive and thermal desorption. This analysis will be facilitated by
the comparison with laboratory measurements and theoretical
computations carried out (also) by members of the SOLIS team (see, for
example, Minissale et al. 2016 and Skouteris et al. 2017).
\begin{table*}
  \caption{Molecules targeted by the SOLIS observations. They will be
    observed with the high
    spectral resolution provided by the Narrow Band correlator (see Table
    \ref{tab:setups}). The third and fourth columns report the upper
    level energy range of the lines in the WideX correlator and their
    number. Last column lists a very brief summary of the formation of
    each molecule. This is not meant to be exhaustive, but just give a
    general overview. Specific SOLIS articles will address the issue
    in more depth. } \label{tab:SOLIS-molecules}
  \begin{tabular}{lccc|l}
    \hline \hline
    \multicolumn{2}{c}{Molecule} & E$_{up}$ (K) & Lines & Role in chemistry \\
    \hline
    Methoxy & CH$_3$O & 4--83 & 5 & Key radical, precursor of dimethyl ether and other iCOMs. \\
                  & & & & Predicted to form either via the reaction CH$_3$OH + OH in the gas (measured\\
                  & & & & rate), or via photolysis of frozen CH$_3$OH on the grain mantles (rate not \\
                  & & & &  measured/known).\\
    \hline
    Methanol & CH$_3$OH & 7--30 & 6 & A “mother” molecule, gas phase precursor of several iCOMs. \\
                  & & & & Formed on the grain surfaces by hydrogenation of frozen CO (measured\\
                  & & & & rate but debated). Not clear how it is released in the gas phase in cold gas,\\
                  & & & & possibly by cosmic rays or reactive desorption, but with uncertain rates.\\
    \hline
    Dimethyl ether &  CH$_3$OCH$_3$ & 19--150 & $\geq$30 & Predicted to form on the grain surfaces via UV/particle irradiation and/or\\
    (DME)     & & & & from radicals on the surfaces, or in the gas phase by the reaction of CH$_3$O\\
                 & & & & and CH$_3$. Its detection in cold environments is difficult to explain within\\
                 & & & & our current understanding of grain-surface radical chemistry, as radicals can not\\
                 & & & &  diffuse at low temperatures. The release from the grains into the gas is problematic as \\
                 & & & & laboratory experiments show that it would rather break the molecule. \\
                 & & & & The rate of the gas-phase radiative association reaction is highly uncertain, that on \\
                 & & & & the grain surfaces is unknown.\\
    \hline
    Methyl formate & HCOOCH$_3$ & 30--550 & $\geq$20 & Predicted to form either on the grain surfaces by reaction of CH$_3$O with HCO\\
    (MF)      & & & & and O (rate known), or in the gas phase from a sequence of reactions starting\\
                & & & & from DME (measured rates). As for DME, the grain-surface formation route is\\
                & & & & challenged by the MF detection in cold environments  (see text).\\
    \hline
    Formamide & NH$_2$HCO & 4--83 & 4 & Initially predicted to form on the grain surfaces by hydrogenation of\\
                & & & & frozen HNCO, laboratory experiments do not confirm this route. Recent\\
                & & & & theoretical computations predict formation  via the gas-phase reaction of\\
                & & & & NH$_2$ and H$_2$CO.\\
    \hline
  \end{tabular}
\end{table*}

The observations of the species listed in Table
\ref{tab:SOLIS-molecules} are obtained via five setups, summarised in
Table \ref{tab:setups}.  In addition to the five molecules in Table
\ref{tab:SOLIS-molecules}, the NOEMA WideX large band correlator
provides us with several other lines and species, including more
iCOMs, which are a precious complement to constrain further the
models.

\begin{table*}
  \caption{List of the frequency setups of SOLIS. The columns
    report the setup number, the spatial resolution in arcsec and au,
    the frequency range in the WideX and Narrow Band correlator, the
    velocity resolution of the latter, the species in Table \ref{tab:SOLIS-molecules}
    targeted with the Narrow Band correlator, and the reached
    rms. Please note that the present article only reports
    observations obtained with setup 1. }\label{tab:setups}
  \begin{tabular}{lccccclc}
    \hline \hline
    Setup & \multicolumn{2}{c}{Spat.Res.} & \multicolumn{2}{c}{Frequency range (GHz)} & Vel.Res.  & Species & rms\\
             &  ($''$)  & (au)                             &   WideX    & Narrow Band                        & (km/s)   &             & (mJy/beam)\\ 
    \hline
    1$^{a,c}$ & $\sim$4 & 300--1000 & 80.80--84.40 & 81.60--82.60 & 0.57 & Methoxy, Formamide & 4--5 \\
    2$^b$ & $\sim$4 & 300--1000  & 80.80--84.40 & 81.60--82.60 & 0.14 & Methoxy, Formamide & 8--9 \\
    3$^a$ & $\sim$4 & 300--1000  & 95.85--99.45 & 96.65--97.65 & 0.48 & Methanol, Dimethyl ether, Methyl formate & 4--5 \\
    4$^b$ & $\sim$4 & 80--200  & 95.85--99.45 & 96.65--97.65 & 0.12 & Methanol, Dimethyl ether, Methyl formate & 8--9 \\
    5$^a$ & $\sim$1 & 80--200   & 204.0--207.6 & 204.8--205.8 & 0.91 & Methanol, Dimethyl ether, Methyl formate & 7--14 \\
    \hline
  \end{tabular}\\
  {\small $^a$ This setup is only used for the ``hot'' sources, namely NGC1333-IRAS4A, CepE, NGC1333-SVS13A, OMC-2 FIR4 and L1157-B1.}\\
  {\small $^b$ This setup is only used for the ``cold'' sources,
    namely L1544 and L1521F.}\\
  {\small $^c$ In the case of OMC-2 FIR4, we only used
    configuration D and, therefore, we reached an
    angular resolution of $\sim$9$''$.5x6.0$''$, while in L1157-B1 and IRAS4A both configurations
    C and D were used and we reached an angular resolution of $\sim$3.5--4$''$. }
\end{table*}
%


\section{Observations and results}\label{sec:observations-results}
\subsection{Observations and analysis}
We report here the first set of observations carried out towards three
SOLIS targets: L1157-B1, OMC-2 FIR4 and NGC1333-IRAS4A. The 
observations took place during the summer-winter 2015 in the setup 1 of Table
\ref{tab:setups}. We used the array in configurations D and C, with
baselines from $\sim 15$ to $\sim 240$ m, providing angular
resolutions going from $\sim 9''.5\times 6''.0$ for the D configuration
only (used for OMC-2 FIR4) to $\sim 4''.0 - 3''.5$ for the combined C+D
configurations (used for L1157-B1 and NGC1333-IRAS4A). The phase
centers and local standard of rest velocities were set to the values
listed in Table \ref{tab:sources}. The primary beam is $\sim 61''$. The
system temperature was between 100 and 200~K in almost all tracks, and
the amount of precipitable water vapour was generally around
10~mm. The calibration of the bandpass and of the absolute flux scale
were performed on the usual NOEMA calibrators, 3C454.3 and MWC349
(when available). The calibration of the gains in phase and amplitude
was performed on strong quasars close to each source (angular distance
$\leq 20^o$). For details on the calibrators used for each individual
source, please see the specific articles listed in the Introduction.

We used both the Narrow Band and the WideX correlators, which provide
us data with different bandwidths and spectral resolutions (see Table
\ref{tab:setups}). The continuum of each source was imaged by
averaging the line-free channels of the WideX correlator
units. Calibration and imaging were performed using the CLIC and
MAPPING softwares of the GILDAS\footnote{The GILDAS software is
  developed at the IRAM and the Observatoire de Grenoble, and is
  available at http://www.iram.fr/IRAMFR/GILDAS.} package using
standard procedures. For OMC-2 FIR4 and NGC1333-IRAS4A, for which a
strong continuum was detected, the continuum image was self-calibrated
(in phase and amplitude), and the solutions were applied to the
lines. This could not be performed for L1157--B1, for which the
continuum was not detected. The analysis of the spectral lines was
performed using standard procedures of the CLASS software, which is
part of the GILDAS package mentioned above. All of the spectral
parameters used to interpret the observations are retrieved from the
Cologne Database for Molecular Spectroscopy (CDMS; M\"{u}ller et
al. 2001, M\"{u}ller et al. 2005).

\subsection{Results}\label{sec:results}
\begin{figure*}
  \centering
  \includegraphics[angle=0,width=19cm]{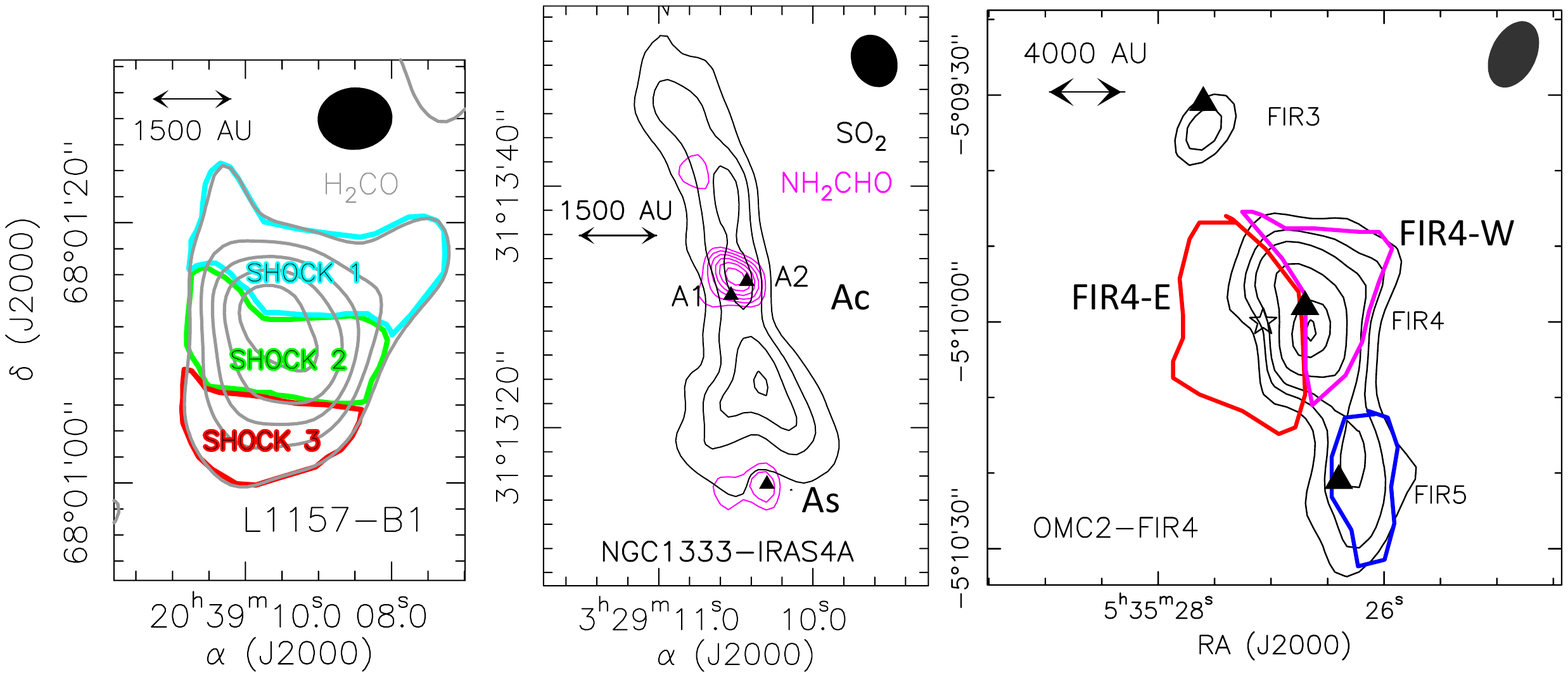}
  \caption{Maps of the three regions observed by SOLIS and reported in
    this article. The synthesised beam are shown at top right of each
    plot. {\it Left panel, L1157-B1:} Contour map of the
    H$_2$CO ($2_{0,2}-1_{0,1}$) line emission (Benedettini et al. 2013)
    overlapped with the three regions 
    identified by the SOLIS observations of the formamide line
    emission (see Tab. \ref{tab:results} and Codella et al. 2017):
    $SHOCK$ 1 is the region with the shock caused by the most recent
    impact of the jet with the B1 cavity wall, while $SHOCK$ 3 is the
    oldest shocked region. The first contour and steps are 3$\sigma$
    (1$\sigma$ = 3.3 mJy beam$^{-1}$ km s$^{-1}$) and 6$\sigma$,
    respectively.  
    {\it Middle panel, IRAS4A:} Contour map of the NH$_2$CHO
    4$_{1,4}-3_{1,3}$ line (magenta contours) 
    overlapped to the SO$_2$ ($8_{1,7}-8_{0,8}$) line emission (black
    contours) obtained with SOLIS observations. The south lobe
    is blue-shifted while the north one is red-shifted. The black
    triangles mark the positions of the two central objects, A1 and A2
    (Looney et al. 2007), called Ac in this article, and the south
    shocked region, As. The 
    first contour and steps are 3$\sigma$ (1$\sigma$ = 
    0.7 mJy beam$^{-1}$ km s$^{-1}$) and 6$\sigma$, respectively.  
    {\it Right panel, OMC-2:} SOLIS continuum contour map at 82 GHz of 
    the OMC-2 region (Fontani et al. 2017). The three black triangles
    show the positions 
    of the three sources in the region, FIR3, FIR4 and FIR5, while the
    white star marks the position of the Herschel/PACS Mid-IR source
    (Furlan et al. 2014). The FIR4 source contains two distinct
    regions based on the HC$_5$N emission: FIR4-E is the eastern
    region (red contour), where the HC$_5$N line is bright, while
    FIR4-W is the western region (magenta contour) with weak HC$_5$N
    line (Fontani et al. 2017). The FIR5 mission extends towards south (blue
    contour). The contour levels of the continuum start from 3$\sigma$
    rms, 0.18mJy/beam, and increase by a factor 2 each time.
  }\label{fig:maps}
\end{figure*}

Figure \ref{fig:maps} shows the maps of the three imaged sources:
OMC-2 FIR4, NGC1333-IRAS4A and L1157-B1. In each map, we marked the
regions which posses the spectra analysed in the following.
Please note that specific articles (see Introduction), are devoted to
each single source. Our goal here is to show the importance of having
interferometric observations to disentangle and reveal the chemical
similarities and differences inside each region within a source and
among the different sources at various scales.

\noindent {\it L1157-B1}: We identified three regions, based on the
analysis presented in Codella et al. (2017): {\it SHOCK} 1, 2 and
3. Briefly, based on the different abundance distribution of
acetaldehyde and formamide, the region {\it SHOCK} 1 contains the gas
most recently shocked by the jet impinging on the B1 cavity wall,
whereas the region {\it SHOCK} 3 marks the oldest shocked gas in B1,
with a difference in time of $\leq$1000 yr (see also Podio et
al. 2016).

\noindent {\it OMC-2 FIR4}: The NOEMA primary beam encompasses three
sources: FIR3, FIR4 and FIR5. In Fontani et al. (2017), we show that
FIR4 consists of two regions, West and East, marked in the following
FIR4-W and FIR4-E respectively. These two regions differ in the
HC$_3$N (9-8)/HC$_5$N (31-30) line emission ratio. Considering the
possible difference in the excitation conditions, the different line
emission ratio translates into a different HC$_3$N/HC$_5$N abundance
ratio and/or gas temperature of the two regions.

\noindent {\it NGC1333-IRAS4A}: We identified two regions. The first
one, marked Ac, is coincident with the well known hot corino
IRAS4A, which consists of two objects, A1 and A2, not resolved by the
present observations (e.g. Taquet et al. 2015; L\'opez-Sepulcre et
al. 2017). The second region, marked as As, is south-east of the hot
corino. It presents weak, but definitively detected, formamide line
emission. Note that As, as in L1157-B1, no continuum is detected there,
so that it is very likely a shock site. Supporting this hypothesis,
the formamide emission is blue-shifted as other tracers of the two
jets emanating from A1 and A2, and the emission region coincides with
the apex of the SO$_2$ emission that probes the jets (Santangelo et
al. 2016).

\begin{figure*}
  \includegraphics[angle=90,width=16cm]{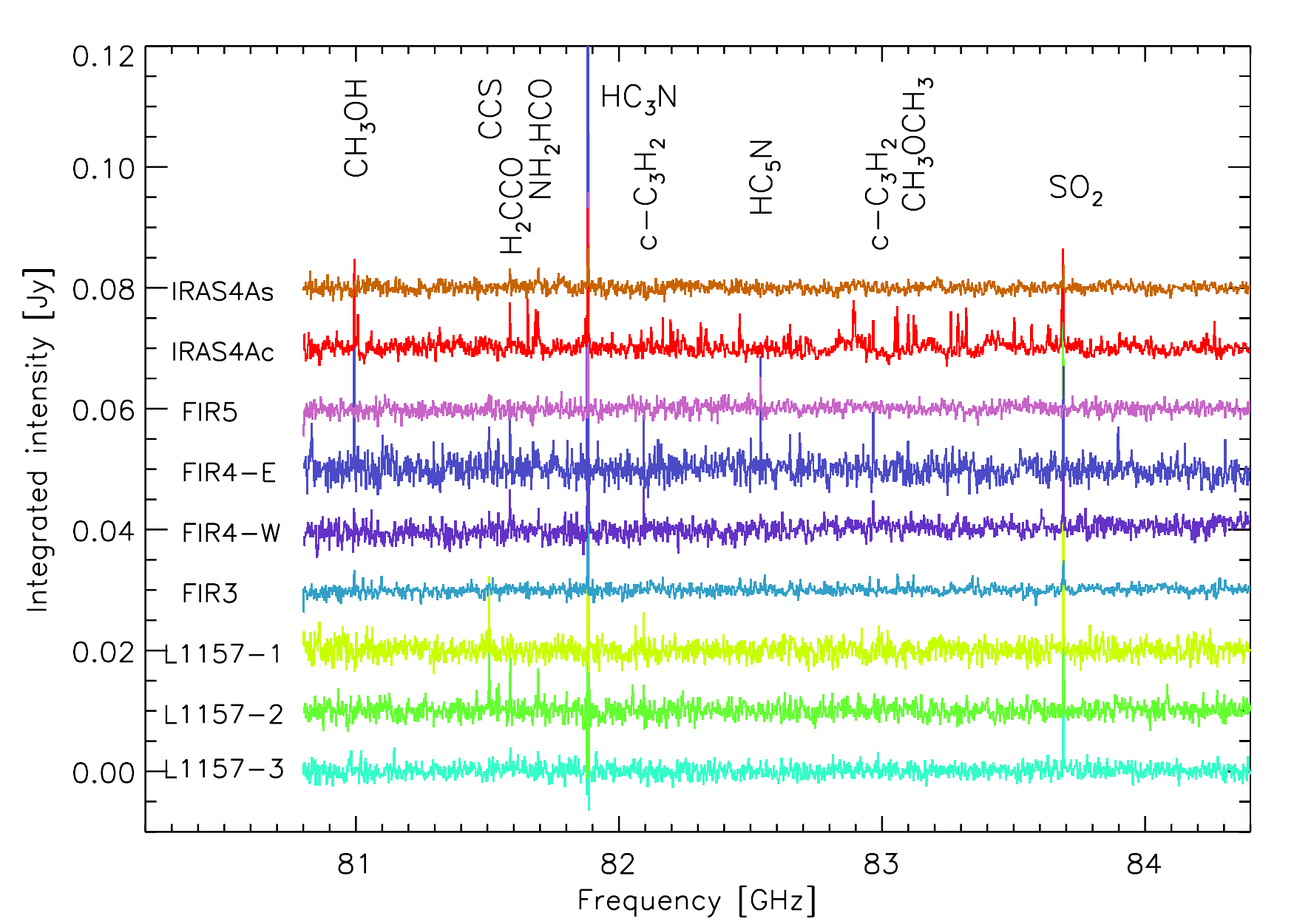}
  \caption{WideX spectra of the eight regions described in the text
    and marked in Figure \ref{fig:maps}, from the bottom: L1157-3 =
    L1157-B1 {\it SHOCK} 3, L1157-2 = L1157-B1 {\it SHOCK} 2, L1157-1
    = L1157-B1 {\it SHOCK} 1, FIR3 = OMC-2 FIR3, FIR4-W = OMC-2 FIR4
    West, FIR4-E = OMC-2 FIR4 East, FIR5 = OMC-2 FIR5, IRAS4Ac =
    IRAS4A hot corino, and IRAS4As = IRAS4A shocked region. The
    integrated intensity is in Jy and the frequency in GHz. An
    arbitrary offset is added to each spectrum. The brightest detected
    lines are labeled at the top of the plot.}\label{fig:widex-whole}
\end{figure*}

\begin{figure*}
  \centering
  \includegraphics[angle=90,width=16cm]{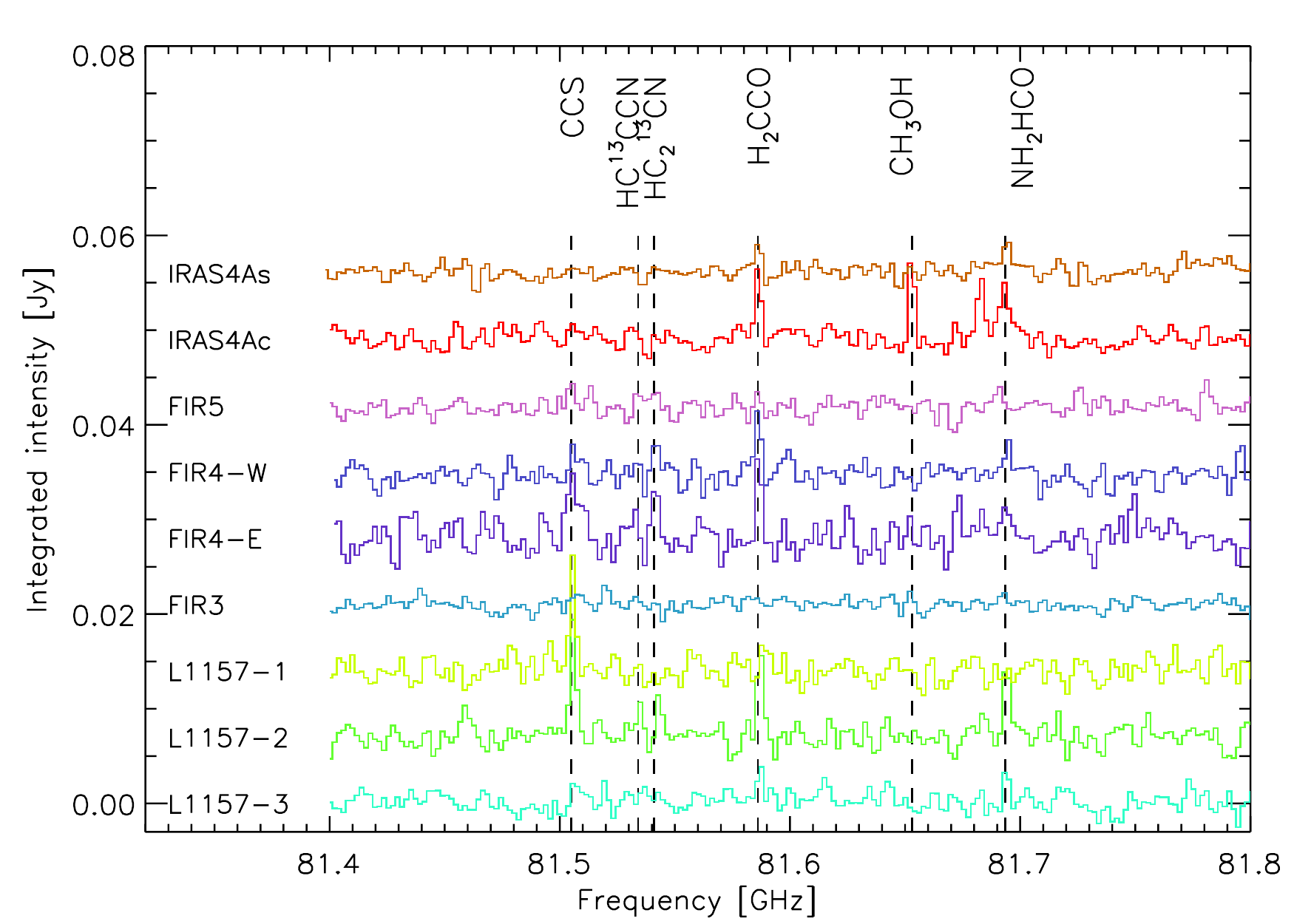}
  \caption{SOLIS setup1 WideX spectra of the eight regions described
    in the text and marked in the figure, following the notation of
    Fig. \ref{fig:widex-whole} and Tab. \ref{tab:results}. The integrated intensity is in Jy and
    the frequency in GHz. The brightest detected lines are labeled at
    the top of the plot.}\label{fig:widex-zoom}
\end{figure*}

The WideX spectra obtained by integrating the emission over the above
regions are shown in Fig. \ref{fig:widex-whole}. A zoom-in around the
81.5 GHz frequency range, which is particularly rich in lines, is
shown in Fig. \ref{fig:widex-zoom}. Table \ref{tab:results} summarises
the detected lines in each source/region and the measured integrated
fluxes. Note that we consider a detection when the line integrated
flux is above 5$\sigma$.
\begin{table*}
  \caption{List of lines detected by SOLIS
    in the 80.80--84.40 GHz frequency range (setup 1 of Table
    \ref{tab:setups}) towards the regions NGC1333-IRAS4A (IRAS4A in
    the header), the OMC-2 FIR4 (OMC-2 in the header) and
    L1157-B1$^a$. ``Y'' means that the line is detected$^{a}$, ``N'' 
    means that the line is undetected. Additional lines are detected
    towards NGC1333-IRAS4A, but we report in this Table only the lines
    detected in at least two sources.}\label{tab:results}
  \begin{tabular}{lcccc|cc|cccc|ccc}
  \hline \hline
Species &Transition & $\nu^{c}$ & $E_{\rm u}^{c}$ & log($A_{\rm ij}$/s$^{-1}$)$^{c}$ &
\multicolumn{2}{c}{IRAS4A$^b$} & \multicolumn{4}{c}{OMC-2$^b$} & \multicolumn{3}{c}{L1157-B1$^b$} \\ 
& & (GHz) & (K) & & As & Ac & FIR5 & FIR4-W & FIR4-E & FIR3 & 1 & 2 & 3\\ 
\hline
CH$_3$OH & 7$_{2,6}-8_{1,7}$ A & 80.99324 & 103 & --5.98 & N & Y & N & N & Y & N & N & N & N \\
CCS &  $6_7-5_6$ & 81.50517 & 15 & --4.61                      & N & N & N & N & Y & N & Y & Y & N \\ 
HC$^{13}$CCN &  9--8 & 81.53411 & 20 & --4.38               & N & N & N & N & Y & N & N & Y & N \\ 
HCC$^{13}$CN &  9--8) & 81.54198 & 20 & --4.38              & N & N & N & N & Y & N & N & Y & N \\ 
H$_2$CCO &  4$_{1,3}-3_{1,2}$ & 81.58623 & 23 & --5.27    & Y & Y & N & N & Y & N & N & Y & N \\
NH$_2$CHO &  4$_{1,4}-3_{1,3}$ & 81.69345 & 13 & --4.43 & Y & Y & N & Y & Y & N & N & Y & Y\\ 
HC$_3$N &  9--8 & 81.88147 & 20 & --4.38                     & Y & Y & Y & Y & Y & Y & Y & Y & Y \\
c-C$_3$H$_2$ &  2$_{0,2}-1_{1,1}$ & 82.09354 & 6 & -4.72 & N & N & N & Y & Y & N & Y & Y & N \\ 
HC$_5$N &  31--30 & 82.53904 & 63 & -4.21                   & N & N & Y & N & Y & N & N & N & N \\ 
c-C$_3$H$_2$ &  3$_{1,2}-3_{0,3}$ & 82.96620 & 16 & --5.00 & N & Y & N & Y & Y & N & N & N & N \\ 
SO$_2$ &  8$_{1,7}-8_{0,8}$ & 83.68809 & 37 & --5.17      & Y & Y & N & Y & Y & Y & Y & Y & Y \\ 
\hline
\end{tabular}
\\
{\small $^a$As reported in the text, lines are considered detected 
  when the flux peak is at least 5 times the rms noise.}\\
{\small $^b$We report the results towards the eight regions described in the
  text: L1157-1 = L1157-B1 {\it SHOCK} 1, L1157-2 = L1157-B1 {\it
    SHOCK} 2, L1157-3 = L1157-B1 {\it SHOCK} 3, 
  FIR3 = OMC-2 FIR3, FIR4-W = OMC-2 FIR4 West, FIR4-E = OMC-2 FIR4 East,
  FIR5 = OMC-2 FIR5, IRAS4Ac = IRAS4A hot corino, and IRAS4As =
  IRAS4A shocked region.}\\
{\small $^c$The Cologne Database for Molecular Spectroscopy
  (CDMS; {\it http://www.astro.uni-koeln.de/cdms/};
  M\"uller et al. 2001, 2005) molecular database was used for
  retrieving the spectroscopic data obtained by Xu \& Lovas (1997),
  Saito et al. (1987), Lovas et al. (1992), Thorwirtht et al. (2000
  and 2001), Fabricant et al. (1977), Brown et al. (1990), Kryvda et
  al. (2009), Spezzano et al. (2012), Bizzocchi et al. (2004), and
  M\"{u}ller et al. (2005).}
\end{table*}

\begin{table*}[bt]
  \caption{Integrated intensity in K km/s of the detected lines
    listed in Table \ref{tab:results}. The last row reports the rms  in K
    towards each region.}\label{tab:fluxes}
 
 \begin{tabular}{lcccccccccc}
    \hline \hline
    Species & Frequency & \multicolumn{9}{c}{Integrated flux (K km/s)}\\
                &    (GHz)        &  IRAS4As &   IRAS4Ac &      FIR5 &    FIR4-W &    FIR4-E &      FIR3 &    L1157-1 &    L1157-2 &    L1157-3\\
    \hline
       CH$_3$OH &    80.993 &           &     0.195 &           &           &     0.305 &     &           &           &          \\
            CCS &    81.505 &           &           &           &           &     0.172 &           &     0.132 &     0.139 &          \\
   HC$^{13}$CCN &    81.534 &           &           &           &           &     0.081 &           &           &     0.045 &          \\
   HCC$^{13}$CN &    81.542 &           &           &           &           &     0.079 &           &           &     0.046 &          \\
        H$_2$CCO &    81.586 &     0.042 &     0.082 &           &     0.077 &     0.113 &           &           &     0.135 &          \\
      NH$_2$CHO &    81.693 &     0.041 &     0.058 &           &     0.039 &     0.053 &           &           &     0.086 &     0.074\\
        HC$_3$N &    81.881 &     0.091 &     0.309 &     0.524 &     2.039 &     4.202 &     0.360 &     2.116 &     3.751 &     1.565\\
   c-C$_3$H$_2$ &    82.093 &           &           &           &     0.080 &     0.102 &           &     0.075 &     0.057 &          \\
        HC$_5$N &    82.539 &           &           &     0.063 &           &     0.177 &           &           &           &          \\
   c-C$_3$H$_2$ &    82.966 &           &     0.070 &           &     0.078 &     0.135 &     &           &           &          \\
         SO$_2$ &    83.688 &     0.061 &     0.385 &           &     0.378 &     0.192 &     0.093 &     0.331 &     1.020 &     0.533\\
    
    rms(K)    &            &     0.001 &     0.002 &     0.001 &     0.010 &     0.005 &     0.001 &     0.005 &     0.008 &     0.008\\
    \hline
\end{tabular}
\end{table*}

\begin{figure*}
  \centering
  \includegraphics[angle=90,width=16cm]{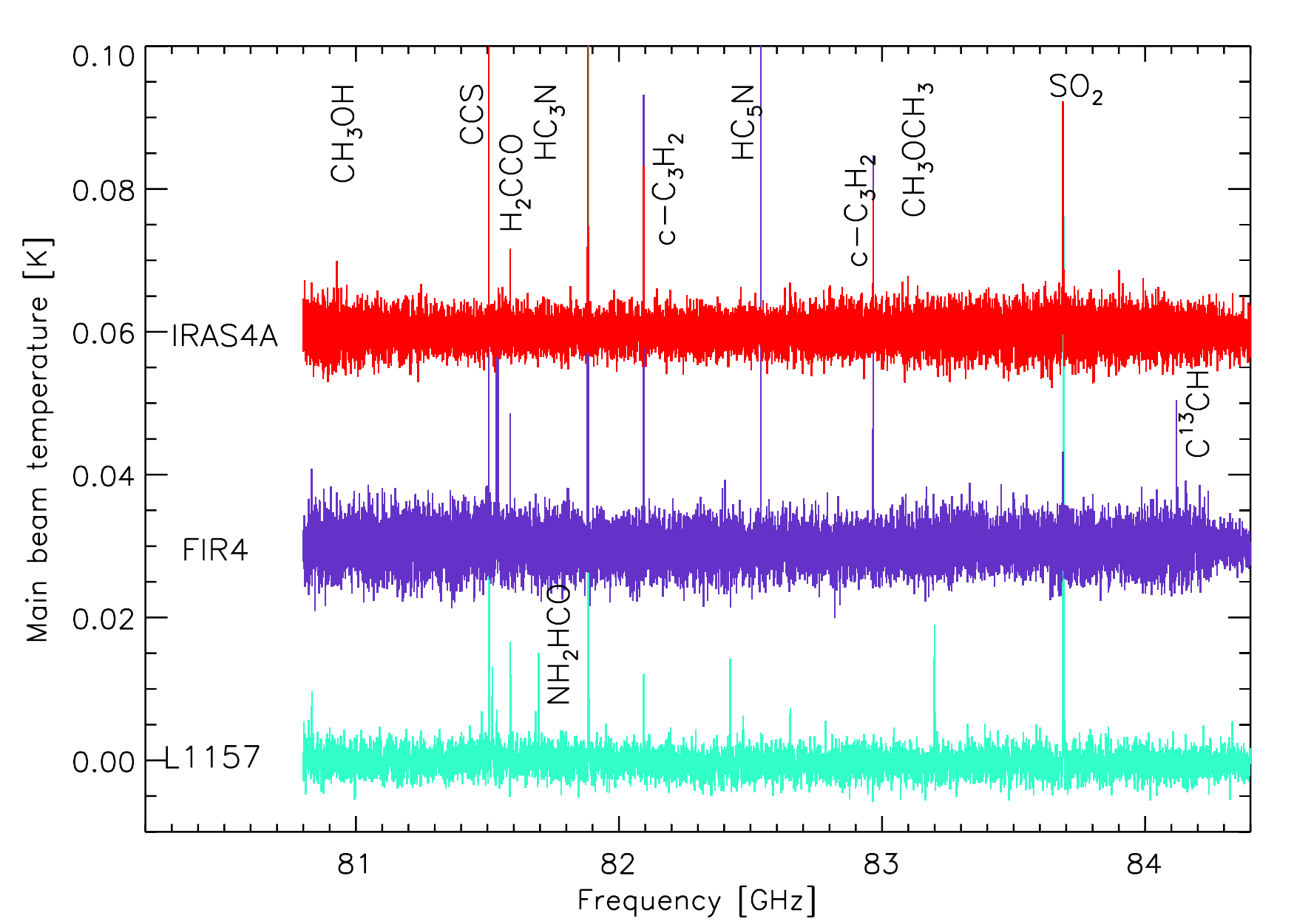}
  \caption{IRAM-30m spectra of the three regions L1157-B1, OMC-2 FIR4
    and IRAS4A in the same frequency range of
    Fig. \ref{fig:widex-whole}. The integrated intensity is in main
    beam temperature (K) and the frequency in GHz. The brightest
    detected lines are labeled at the top and bottom of the plot. The data were
    taken in the framework of the Large Project ASAI (Lefloch et
    al. 2017b).}\label{fig:asai}
\end{figure*}

Finally, for comparison, in Figure \ref{fig:asai} we show the IRAM-30m
spectra obtained towards the three sources in the same frequency range
of Fig. \ref{fig:widex-whole}. At these frequencies, the main beam of
IRAM-30m is $\sim 30''$, namely about three to ten times larger than
the NOEMA synthesised beams. Note that the IRAM-30m spectra were
obtained within the Large Project ASAI\footnote{Astrochemical Surveys
  At IRAM: http://www.oan.es/asai/} described in Lefloch et al. (2017b).


\section{Discussion}\label{sec:discussion}

We organised the discussion in three parts. The first one discusses the
overall similarities and diversity between the three sources and the
regions composing them, while a second part discusses the three
sources separately. A third section discusses the detection and
non-detection of the two SOLIS target molecules of setup 1: formamide
and methoxy.

\subsection{Overall sources similarity and diversity}

\subsubsection{Large (3000-10000 au) scale line emission} 
The comparison between the SOLIS (Fig. \ref{fig:widex-whole}) and ASAI
(Fig. \ref{fig:asai}) spectra provides a straightforward information
on the emission at the 3000-10000 au scales probed by the ASAI
observations and filtered-out by the SOLIS ones. For example, the CCS
line is very bright in the ASAI spectra, and weak (L1157-B1 {\it
  SHOCK} 1 and 2, and FIR4-E), if not absent (IRAS4Ac and FIR5), in
the SOLIS ones, a clear indication that CCS is abundant in the
extended molecular clouds to which L1157-B1, OMC-2 FIR4 and
NGC1333-IRAS4A belong. The same can be said of the H$_2$CCO,
c-C$_3$H$_2$ and HC$_5$N lines, which are prominent in the ASAI
spectra and weak or absent in the SOLIS ones. All these species are
small hydrocarbons or unsaturated C-bearing chains particularly
abundant in the skin of the parental clouds (e.g. Spezzano et
al. 2017). Two dedicated SOLIS articles discuss the details regarding
the cyanopolyynes (Fontani et al. 2017) and c-C$_3$H$_2$ (Favre et
al. 2017) towards OMC-2 FIR4.

Note that the ASAI spectra are relatively similar in OMC-2 FIR4 and
NGC1333-IRAS4A, with the biggest exception being the HC$_5$N line,
totally absent in NGC1333-IRAS4A and very bright in OMC-2 FIR4. In
Fontani et al. (2017), we argue that this is due to the anomalously large
energetic ($\geq 10$ MeV) particle irradiation of the OMC-2 FIR4
region, first revealed by {\em Herschel} observations (Ceccarelli et
al. 2014). On the other hand, the similarity suggests the presence of
an embedded hot corino in OMC-2 FIR4. 

The shocked region L1157-B1 shows a relatively different line spectrum
with respect to both OMC-2 FIR4 and NGC1333-IRAS4A, with bright lines
of formamide (Codella et al. 2017) and SO$_2$ (Feng et
al. in prep.). This difference with respect to OMC-2 FIR4 is less
surprising, as the single-dish ASAI spectrum in the latter is likely
dominated by a dense and cold gas highly irradiated by energetic
particles (Ceccarelli et al. 2014; Fontani et al. 2017), namely it is
likely due to the very different environment of OMC-2 FIR4 and
L1157-B1. A bit more surprising is the difference between L1157-B1 and
NGC1333-IRAS4, as in both sources the emission presumably (also)
originates from species directly or indirectly injected by the grain
mantles. This difference is probably due to either a different
chemical timescale or a different initial grain mantle composition.

\subsubsection{Chemical diversity at 300-1000 au scale} 
The sources and the regions surrounding them appear particularly
different at the 300-1000 au scale. The spectra of
Figs. \ref{fig:widex-whole} and \ref{fig:widex-zoom} clearly identify
either different zones of line emission or different chemical
composition across each target source.  

\begin{figure*}
  \centering
  \plotone{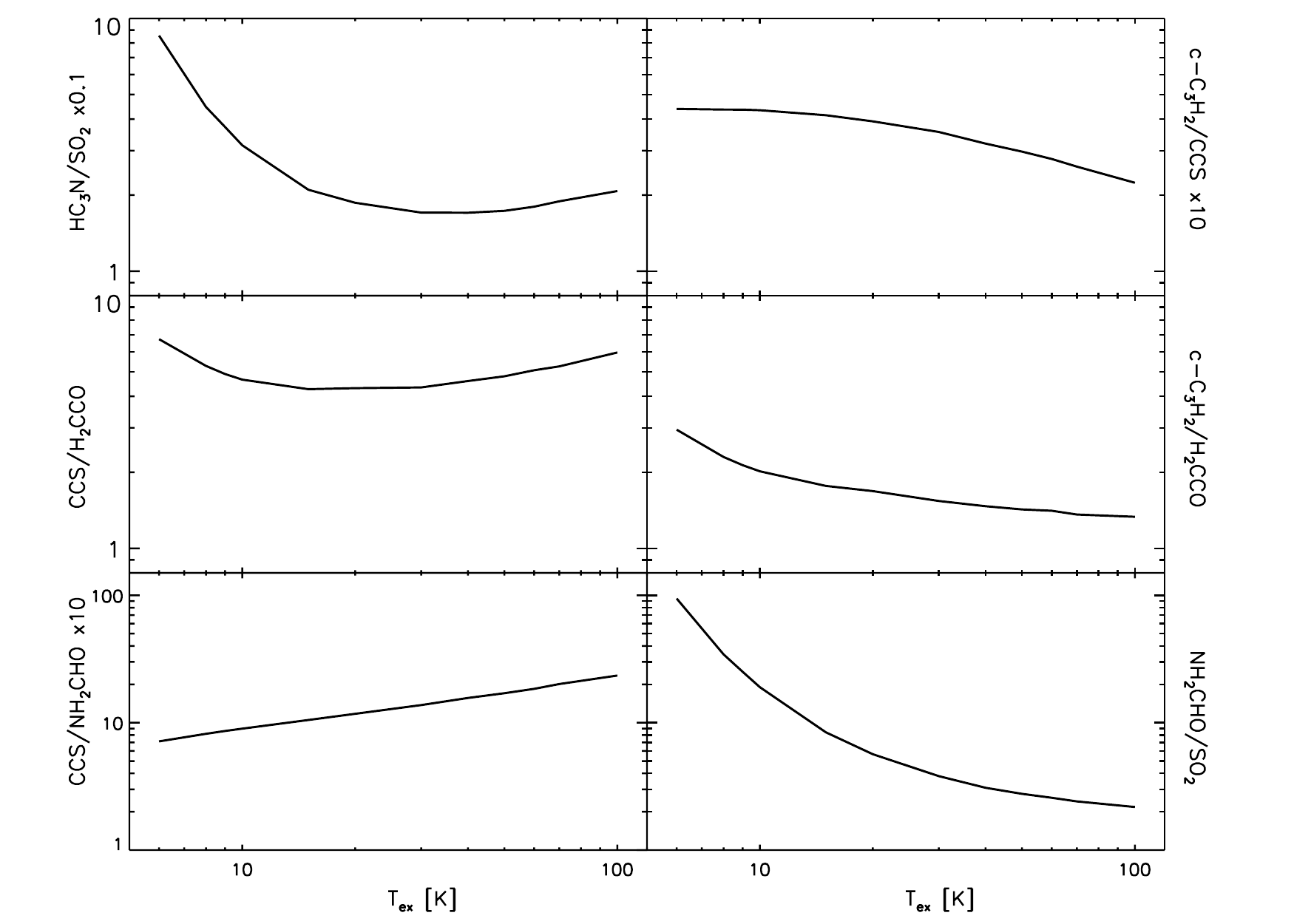}
\caption{Line ratios as a function of the excitation temperature
  T$_{ex}$: HC$_3$N 9--8 / SO$_2$ 8$_{1,7}-8_{0,8}$ multiplied by a
  factor 0.1 (left top), CCS $6_7-5_6$ / H$_2$CCO 4$_{1,3}-3_{1,2}$
  (left middle), CCS $6_7-5_6$ / NH$_2$CHO 4$_{1,4}-3_{1,3}$ (left
  bottom), c-C$_3$H$_2$ 3$_{1,2}-3_{0,3}$ / CCS $6_7-5_6$ multiplied
  by a factor 10 (right top), c-C$_3$H$_2$ 3$_{1,2}-3_{0,3}$ /
  H$_2$CCO 4$_{1,3}-3_{1,2}$ (right middle), and NH$_2$CHO
  4$_{1,4}-3_{1,3}$ / SO$_2$ 8$_{1,7}-8_{0,8}$ (right bottom). The
  line ratios were computed assuming the same column density for all
  species, LTE populated and optically thin lines.}\label{fig:line-ratios}
\end{figure*}

All the 80.8-84.3 GHz spectra are dominated by the HC$_3$N (9-8) line
and, to a lesser extent, the SO$_2$ ($8_{1,7}-8_{0,8}$) line (see
Table \ref{tab:fluxes}). Yet, the relative intensity with respect to
other lines varies from region to region. In particular, the HC$_3$N
(9-8) to SO$_2$ ($8_{1,7}-8_{0,8}$) intensity ratio varies from about
unity in NGC1333-IRAS4As to 20 in OMC-2 FIR4-E. Such a large
difference is unlikely due to excitation effects, for the two lines
have rather similar upper level energies, so it probably reflects a
real chemical difference. Figure \ref{fig:line-ratios} shows the
  predicted line ratio as a function of the excitation temperature,
  assuming LTE and optically thin lines. The figure shows that this
  ratio varies by a factor of about four between 6 and 100 K.
Therefore, NGC1333-IRAS4As appears to be a region enriched in SO$_2$
compared to OMC-2 FIR4-E or, alternatively, poorer in
HC$_3$N. L1157-B1 lies in between, with a HC$_3$N (9-8) to SO$_2$
($8_{1,7}-8_{0,8}$) line ratio of 3--6. This is indeed consistent
  with previous observations that show an enhancement in shocked
  regions of SO$_2$ but not of HC$_3$N (e.g. Bachiller \& Perez
  Gutierrez 1997; Benedettini et al. 2013).

Even more marked is the difference in intensity of weaker lines.
Interestingly, the CCS ($6_7-5_6$) line is relatively bright in
L1157-B1 {\it SHOCKS} 1 and 2, and in OMC-2 FIR4-E, while it is
undetected in the other sources/regions. On the contrary, the H$_2$CCO
($4_{1,3}-3_{1,2}$) line is bright in L1157-B1 {\it SHOCK} 2 and OMC-2
FIR4-E, weak in NGC1333-IRAS4Ac and OMC-2 FIR4-W, and undetected
elsewhere. Also in this case, the largest differences are likely
caused by a real difference in chemical composition rather than
excitation conditions, as the two lines have, again, similar upper
level energies. This is quantitatively shown in Figure
  \ref{fig:line-ratios}: the CCS ($6_7-5_6$) over H$_2$CCO
  ($4_{1,3}-3_{1,2}$) line ratio is practically constant for an
  excitation temperature varying between 6 and 100 K.

More relevant for the present work and the goal of SOLIS, lines from
formamide are variably present in the sources/regions. For example,
the NH$_2$CHO ($4_{1,4}-3_{1,3}$) line is detected in L1157-B1 {\it
  SHOCKS} 2 and 3, NGC1333-IRAS4A hot corino and shock, tentatively in
OMC-2 FIR4 West and East, and undetected in OMC-2 FIR3 and FIR5
(Fig. \ref{fig:widex-zoom}). When considering the ratio between the
formamide and CCS lines, which have very similar upper level energies
and Einstein coefficients, the chemical differentiation is
particularly evident with CCS/NH$_2$CHO varying from 1.5--3 in OMC-2
FIR4-E and L1157-B1 {\it SHOCK} 2 to less than unity in the other
regions where NH$_2$CHO is detected. Figure \ref{fig:line-ratios}
  quantifies how different the excitation conditions should be to
  explain the observed difference in the CCS/NH$_2$CHO line ratio. For
  excitation temperatures between 10 and 80 K, the line ratio varies
  by no more than a factor two. Therefore, dedicated studies of the
  various sources/regions coupled with chemical modelling, as in
  Codella et al. (2017), are needed to shed light on the processes
  behind the observed line intensity differences among the various
  sources.

In summary, the distribution and relative abundances of (complex)
organic molecules change significantly among the different sources,
and this diversity is already clear at 300-3000 au scales.

\subsection{Differences within the sources}

\subsubsection{NGC1333-IRAS4A} 

The spectra of the two studied regions, the hot corino and the shocked
region, are definitively very different (Fig. \ref{fig:maps}). The hot
corino shows the typical lines of iCOMs (not reported in Tables
\ref{tab:results} and \ref{tab:fluxes} because they are many but not
relevant in the context of this article: the entire list will be
the subject of a forthcoming publication), such as formamide, dimethyl
ether and methanol, whereas the shocked region possesses only two
(relatively) bright lines characteristic of hot corinos, the NH$_2$CHO
(4$_{1,4}-3_{1,3}$) and SO$_2$ ($8_{1,7}-8_{0,8}$) lines. The
distribution of the SO$_2$ line intensity is shown in
Fig. \ref{fig:maps}. The line clearly maps the contour of the
outflowing/shocked gas. Particularly intriguing is the formamide
detection towards the shocked region, in a position which is at the
border of the region where SO$_2$ emits. Note that SO$_2$ traces
fairly well the two jets emitted by A1 and A2, observed in SiO by
Santangelo et al. (2015) and here unresolved. As in the case of
L1157-B1 (Codella et al. 2017; Feng et al. 2017), the SO$_2$ versus
formamide spatial segregation could be caused by a time effect, in
addition to the release in the gas phase of the formamide precursors,
namely formaldehyde and amidogen. It is worth mentioning that methanol
is, rather surprisingly, not detected towards the shocked region,
where formamide is present (Table \ref{tab:fluxes}).  In addition, the
shocked region emits weak lines of H$_2$CCO and HC$_3$N (9-8), also
detected in the hot corino. A forthcoming article will report on a
detailed analysis of this jet plus shocked region (L\'opez-Sepulcre et
al. in prep.).

\subsubsection{OMC-2 FIR4} 

The three objects in the NOEMA primary beam, FIR3, FIR4 and FIR5, show
clear and remarkable differences. The difference in the cyanopolyynes
and c-C$_3$H$_2$ are discussed in Fontani et al. (2017) and Favre et
al. (in prep.), respectively. Briefly, cyanopolyynes probe the presence of
a source of energetic ($\geq10$ Mev) particles irradiating the eastern part of
OMC-2 FIR4 and FIR5. c-C$_3$H$_2$ is, instead, a good thermometer of
the region, which shows a rather uniform temperature with a possible
small increasing gradient toward East, as also found with the low
angular resolution temperature map from NH$_3$ (Friesen \& Pineda et
al. 2017).  This is also consistent with the fact that the region is
irradiated by energetic particles at/from East. 

In addition, the spectra in Figs.  \ref{fig:widex-whole} and
\ref{fig:widex-zoom} indicate the presence of a high lying
(E$_{up}$=103 K) line of CH$_3$OH, typical of hot corinos, only
towards FIR4-E. The SO$_2$ line is bright towards FIR4-W and FIR4-E,
and it is detected in FIR3 but not in FIR5 (Table
\ref{tab:fluxes}). Since SO$_2$ is often also a hot corino species or,
more generally, a warm gas indicator, this would suggest that FIR3
harbours a young protostar, maybe a hot corino, while FIR5 is a cold
object. Similarly, one or more hot corinos might be present in the
FIR4 region.  Towards FIR4, the eastern part (the one irradiated by
energetic particles) is the region with the richest spectrum,
with all of the lines listed in Table \ref{tab:fluxes} detected. It is
likely that one or more hot cores/corinos are embedded, therefore, in
the East region of FIR4. In the western part, only H$_2$CCO and
c-C$_3$H$_2$ are detected.  In general, they are species that are
abundant in the exterior of molecular clouds, where residual UV
photons produce a (small) fraction of neutral carbon (e.g. Spezzano et
al. 2016).

We do not find any evidence that FIR4 North is affected by the outflow
from FIR3, as suggested by Shimajiri et al. (2015): nor an increase of
the gas temperature or broad band emission at North are detected in
the SOLIS observations. Since the Shimajiri et al. observations were
carried out with single-dish telescopes with a beam of $\sim20''$, it
is possible that the emission that they observed is contaminated by
the hot corinos in the East FIR4 region.

Finally, a weak formamide line is present in both FIR4-E and
FIR4-W. To confirm its detection, additional SOLIS observations are
underway. Finally, incoming higher angular resolution SOLIS
observations will definitively establish whether hot corinos or shocks
or both are present and where in the region (Neri et al. in prep.).

\subsubsection{L1157-B1} 

L1157-B1 can be separated into three zones when the relative abundance
of acetaldehyde and formamide is considered (Codella et al. 2017). The
three zones identify three different shocks caused by the episodic
ejections and the resulting jet impacting the cavity wall excavated by
the outflow from the central star. The three shocks have been created
at different times, within a lapse of time of about 1000 yr and with
the one in the north being the youngest one (Podio et al. 2016). The
difference between the three zones is also evident looking at the CCS,
H$_2$CCO and c-C$_3$H$_2$ lines. The CCS and c-C$_3$H$_2$ lines are
bright in the north and central zones, but absent in the south,
whereas H$_2$CCO is only bright in the central zone. This diversity in
the CCS, c-C$_3$H$_2$ and H$_2$CCO line distribution cannot be
attributed to sensitivity, but must reflect a difference in the
excitation conditions or chemical composition of the three
regions. The upper level energies of the CCS, H$_2$CCO and
c-C$_3$H$_2$ (3$_{1,2}-3_{0,3}$) transitions are 15, 22 and 16 K,
whereas the Einstein coefficients are 2.5, 0.5 and 1.0
$\times 10^{-5}$ s$^{-1}$, respectively, so that different excitation
conditions seem unlikely. The plots of these line ratios,
  reported in Figure \ref{fig:line-ratios}, support this
  interpretation.  Based on the analysis of formamide, the line of
which is detected only towards {\it SHOCKs} 2 and 3, we suggest that
the differentiation of CCS, H$_2$CCO and c-C$_3$H$_2$ is also linked
to the evolution of the gas chemical composition on a short time scale
($\sim 1000$ yr; Codella et al. 2017).

\subsection{Methoxy}

\begin{figure}
  \centering
  \plotone{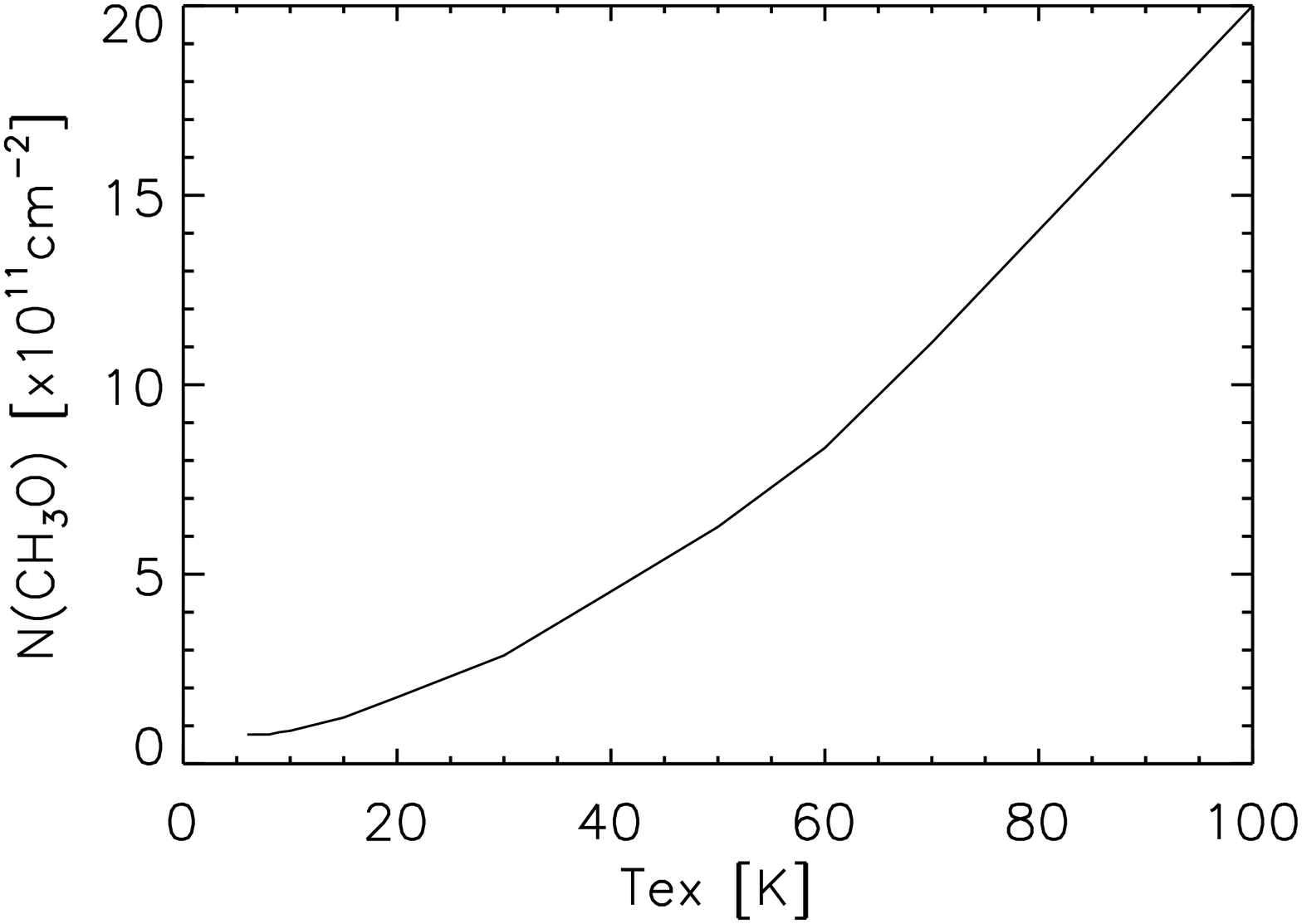}
  \caption{Upper limit to the methoxy column density as a function of
    the excitation temperature, derived assuming a rms of 1 mK, a
    line width of 2 km/s.}\label{fig:methoxy}
\end{figure}

Setup 1 was selected in order to have lines from formamide and methoxy
in the Narrow Band correlator (Table \ref{tab:setups}). While
formamide is detected in all three sources (but not in all regions of
each source: Table \ref{tab:results}), methoxy is undetected in all of
them.
The upper limit to the methoxy column density of each source depends
on the exact physical conditions. However, to have an order of
magnitude, adopting an rms of 1 mK (Table \ref{tab:fluxes}) and a line
width of 2 km/s, the upper limit to the methoxy column density varies
from $8\times10^{10}$ to $2\times10^{12}$ cm$^{-2}$ for an excitation
temperature from 6 to 100 K (Fig. \ref{fig:methoxy})\footnote{The
  upper limit does not apply to extended emission filtered out by the
  NOEMA interferometer.}. This provides a very approximate upper limit
to the methoxy abundance in OMC-2 FIR4 and L1157-B1 of
$2\times10^{-12}$ and $3\times10^{-10}$, assuming an excitation
temperature of 30 and 50 K and a H$_2$ column density of
$2\times10^{23}$ and $2\times10^{21}$ cm$^{-2}$ (Lefloch et al. 2012;
Fontani et al. 2017), respectively. The upper limit to the methoxy
abundance in IRAS4A is approximately $1\times10^{-11}$, assuming an
excitation temperature of 70 K, a H$_2$ column density of
$1\times10^{24}$ cm$^{-2}$ and a hot corino size of $1''.2$
(L\'opez-Sepulcre et al. 2017).

Single-dish observations of cold prestellar objects detected methoxy
so far only towards a handful of cold prestellar cores (Cernicharo et
al. 2012; Jim\'enez-Serra et al. 2016; Bacmann \& Faure 2016). The
measured column densities are about $10^{12}$ cm$^{-2}$. The
derivation of the abundance from these single-dish observations is
rather tricky as it depends on where the observed methoxy line
emission originates (both for the correction of the filling factor and
the adopted H$_2$ column density). Vastel et al. (2014) claimed that
iCOMs emission arises at the border of the prestellar
condensations. Jim\'enez-Serra et al. (2016) indeed showed that
methoxy is about 10 times more abundant in the border of the
prestellar core L1544 than in its interior. Specifically,
Jim\'enez-Serra et al. measured a methoxy abundance equal to
$\sim 3\times10^{-11}$ and an upper limit of $5\times10^{-12}$ at the
border and in the interior of L1544 and are relatively well reproduced
(within a factor of 3) by Vasyunin et al. (2017). These values are
close to the upper limits computed above from the present SOLIS
observations. A forthcoming SOLIS article (Dulieu et al. in prep) will
analyse the chemistry of methoxy using a more appropriate and
dedicated modeling of the SOLIS protostellar sources.

\section{Conclusions}\label{sec:conclusions}
Some interstellar Complex Organic Molecules, in this work called
iCOMs, might have had a role in the emergence of life on Earth,
constituting the smallest bricks to build up biotic macromolecules.  A
few tens of iCOMs are detected so far towards one or two massive star
forming regions, SgrB2 and Orion KL. In Solar-like star forming
regions the number is reduced to a bit more than a dozen, mostly
detected in one source, IRAS16293-2422 (see Introduction), and only a
few of them in other sources (e.g. L\'opez-Sepulcre et al. 2017).
Given the paucity of sources where the species are detected, it is
difficult to assess how the environment affects the presence and
abundance of iCOMs. In turn, this lack of information severely hampers
our understanding on how iCOMs form in the ISM. From a theoretical
point of view, two major routes of formation are possible: on the
grain surfaces or in the gas. More likely, a combination of the two is
at work, but the ``who does what'' is still largely debated.

The advent of new powerful facilities, like IRAM-NOEMA and ALMA, is
opening a new era in the studies of iCOMs, especially toward
Solar-like star forming regions, the ones with the highest potential
in terms of molecules having a possible role in the emergence of
terrestrial life.  The project PILS ({\it Protostellar Interferometric
  Line Survey}: J\o rgensen et al. 2016) is already providing a new
census of iCOMs towards IRAS16293-2422, which improves the old one by
TIMASSS ({\it The IRAS16293-2422 millimeter and submillimeter spectral
  survey}: Caux et al. 2011). In this article, we present a new
IRAM-NOEMA Large Program called SOLIS ({\it Seeds Of Life In Space}),
the goal of which is to obtain the spatial distribution of selected
iCOMs in half a dozen Solar-like star forming regions. The SOLIS
images have spatial resolutions of 1$''$--5$''$, to allow studies
on scales of 100--5000 au. This will allow us to put constraints on
the formation routes of iCOMs, as well as on the mechanisms that
inject them from the grain mantles to the gas, where they are
detected, and the role of UV illumination.

Here we present the SOLIS project and a summary of the first results
obtained towards three sources: NGC1333-IRAS4A, OMC-2 FIR4 and
L1157-B1. They show the potential of obtaining interferometric
images of iCOMs. Comparing the SOLIS and ASAI spectra we can
disentangle the molecular lines emitted in the large-scale envelopes
and molecular clouds harbouring the sources from those in the immediate
vicinity of where the action takes place, namely hot corinos and
shocks.

Also, the SOLIS small scale spectra show a clear differentiation in
the three studied sources, caused by the different physical conditions
in the three different environments. Particularly spectacular is the
difference in the abundance distribution of two cyanopolyynes, HC$_3$N
and HC$_5$N. This is likely caused by a difference in the rate of
energetic ($\geq$ 10 MeV) particles irradiating the various studied
sources, as discussed in detail by Fontani et al. (2017). A similar
conclusion is reached by Favre et al. (2017) from the analysis of the
c-C$_3$H$_2$ emission.

In the setup reported in this study, the SOLIS target iCOM is
formamide. It is firmly detected in the hot corino of NGC1333-IRAS4A,
and, for the first time, in the shocked regions of L1157-B1 and
NGC1333-IRAS4A, created by their respective violent ejections of
material. In OMC-2 FIR4, there is a tentative detection, which needs
to be confirmed by forthcoming SOLIS formamide observations. The
analysis of the distribution of the formamide abundance provides
strong constraints on the formation route of this species. This is
discussed in detail in Codella et al. (2017), who concluded that the
gas-phase reaction proposed by Kahane et al. (2013) and studied by
Barone et al. (2015), Vazart et al. (2016) and Skouteris et al. (2017)
reproduces the SOLIS observations very well.

The other molecule targeted in the setup presented here is
methoxy. This molecules is not detected in any of the three sources,
leading to upper limits to its abundance of $\sim
10^{-12}$--$10^{-11}$, namely the same order of magnitude of that
measured in cold prestellar cores (e.g. Jim\'enez-Serra et al. 2016). A
forthcoming article will exploit this information in detail to set
constraints on the methoxy chemistry.

Further studies based on SOLIS as the ones mentioned here will put
additional strong constraints to the formation and destruction routes
of formamide and, hopefully, also of more iCOMs, particularly those
targeted by SOLIS.

\section{Acknowledgements}
We acknowledge the funding from the European
Research Council (ERC), projects PALs (contract 320620) and DOC
(contract 741002).  
This work was supported by the French program “Physique et Chimie du
Milieu Interstellaire” (PCMI) funded by the Conseil National de la
Recherche Scientifique (CNRS) and Centre National d’Etudes Spatiales
(CNES) 
and by the Italian Ministero dell'Istruzione, Universit\'a e
Ricerca through the grant Progetti Premiali 2012 - iALMA (CUP
C52I13000140001).
 Partial salary support for A. Pon was provided by a Canadian
 Institute for Theoretical Astrophysics (CITA)  National Fellowship.
 I.J.-S. and D. Q. acknowledge the financial support received from the
 STFC through an Ernest Rutherford Fellowship and Grant (proposal
 numbers ST/L004801 and ST/M004139).

{}

\end{document}